# Creating and Maintaining Filipino and Japanese Students' Social Capital with Facebook

**Mayumi Kubota**
Kansai University, Japan
mkubota@res.kutc.kansai-u.ac.jp

**Julius G. Garcia**
Technological University of the Philippines, Philippines
julius.tim.garcia@gmail.com

*This study investigated perceptions and patterns of Facebook (FB) use among Filipino and Japanese undergraduate students and the relationship of these factors to creating and maintaining students' social capital, international posture, and willingness to communicate (WTC). The survey of undergraduate students was conducted online and 483 valid responses were obtained. Data revealed the characteristic uses of FB by Filipino and Japanese undergraduate students. An interrelation model among six factors (International Posture, WTC, Perception, Bridging, Bonding, and Utilization) for Filipino students showed the importance of utilization or FB usage for bridging social capital and bonding social capital. For Japanese students, bonding social capital mediated between utilization or FB usage and bridging. Bridging social capital was established only through bonding social capital. Thus, unless Japanese students are close enough to their FB friends, they do not construct new relationships on FB that will influence Filipino students at the process of "virtual internationalization" for future.*

## Introduction

"Global Human Development" has become a keyword in Japanese higher education's drive to improve Japan's global competitiveness and enhance ties between nations, as well as overcoming younger Japanese's "inward tendency." The Ministry of Education, Culture, Sports, Science and Technology (MEXT) Japan cultivates global humans who can positively meet challenges and succeed globally. In relation to this trend, service learning, study-abroad programs, and international volunteer work have gained more attention, while Kansai University has joined the Center for Collaborative Online International Learning (COIL) network. The main purpose of COIL is to teach students' their target language and cultivate intercultural communication competence through online intercultural exchange among the universities in the world. O'Dowd and Lewis (2016) emphasize the merit of online intercultural exchange (OIE) as a "virtual internationalization," which is "a term that describes the use of OIE to bring together internationally dispersed classes to carry out academic cooperation for mutual benefit" (p. IV). However, most practice is conducted within the class schedule or curriculum of each university. Although many merits have been found, most OIE has occurred in Europe and the U.S., and is "disappointingly small" in Asian countries (O'Dowd & Lewis, 2016, p. 25). In addition, the "virtual internationalization" created through OIE, as proposed by O'Dowd and Lewis (2016), is limited to educational practices, because learning designs must encompass assessment and evaluation. Once students interact through OIE, it is possible for them to freely maintain friendships by using social network sites (SNS), even after their courses together conclude. However, not much attention is given to the possibility of informal learning therein.

Thus, to utilize FB in education this study investigated perception and patterns of Facebook (FB) among Filipino and Japanese undergraduate students and the relationship of these factors in creating and maintaining students' social capital.

## Review of Related Literature

### FB in the Philippines and Japan

Among SNS, FB is the most popular with over one billion monthly users (Internet World Statistics, 2016). Among Asian countries as of 2016, the first-place user was Indonesia (79,000,000), in second place was the Philippines (48,000,000), and Japan was fifth place (24,000,000) (Member, 2016). Regarding the ratio of FB users among internet users, the Philippines were at 102% (first place), Indonesia 101% (second place), and Japan 21% (11th place) (Member, 2016). This means that in the Philippines, all internet users use FB, while in Japan people select FB from diverse SNS sites, such as LINE, YouTube, and Instagram. Thus, Japanese FB usage is limited, another reason being the requirement of using one's real name. Ishii (2013) conducted a comparative cultural study on SNS use by focusing on motivation and self-disclosures, and found that Japanese users had the least self-disclosure, and that their number of friends was the least among countries such as the U.S.A., China, and Taiwan. Thus, the Japanese have unique behaviors regarding the use of FB.

However, FB is so popular in the Philippines partially because of the different orientation of FB's introduction in the country (Kubota, 2016). On March 20, 2015, a telecommunications company partnered with FB introduced the Internet.org app that would give free access to 24 mobile sites in the fields of education, health, jobs, communication, information, and news. The Philippines is the seventh country with an app that offers free mobile Internet services (PhilStar) and the first in Southeast Asia. Subscribers may also use FB and FB Messenger through the app free of charge.

Therefore, although Japanese FB users are few, FB is the easiest way for Japanese and Filipino students to interact with each other after study tour programs in the Philippines. Kubota (2016) investigated the FB contents of 11 Filipino students and 13 Japanese students during and after a 10-day study tour in the Philippines. Analysis of posting and tagging and interview of two Japanese students revealed the specific usages of FB by Filipino and Japanese students. It seems that the difficulty to continuing relationships between Filipino and Japanese students is not only English proficiency but also their behaviors regarding FB, discussed in terms of "pictures posted," "functions of Timeline," and "number of Facebook friends."

FB is utilized to communicate with family, friends, and even strangers. Indeed, FB has created varied modes for young adult users in particular to interact in both personal and public spheres. FB users include college-aged individuals who dedicate an average of approximately 30 minutes a day accessing and integrating it into their daily lives (Pempek, Yermolayeva, & Calvert, 2009). However, inquiry into how FB provides social benefits for its college-aged users in the form of social capital is lacking. This specific demographic is a major concern among educators and policy makers, since these students eventually will become the workforce after they leave campus.

**Social Capital**

Social capital refers to the links and shared social values and understandings that enable individuals and groups to trust each other and work together. It describes the ability of individuals or groups to access resources embedded in their social network (Bourdieu, 1986; Coleman, 1988). Specifically, Bourdieu (1986) identifies three categories of capital; they are economic capital, cultural capital, and social capital. Economic capital is "immediately and directly convertible into money and may be institutionalized in the form of property rights," while cultural capital "is convertible, on certain conditions, into economic capital and may be institutionalized in the form of educational qualifications," and social capital is "made up of social obligations ("connections") and "is convertible, in certain conditions, into economic capital and may be institutionalized in the form of a title of nobility" (p. 243). The more individual cultural and economic capital increase, the more social capital increases, and the more social capital increases, the more economic and cultural capital increase. Since this relationship happens in a cycle, social classifications tend to become rigid.

In sociology and political science, Putnam (1995) defines social capital as "features of social life—networks, norms, and trust—that enable participants to act together more effectively to pursue shared objectives" (p. 664-665). Putnam (2000) identifies two forms of social capital, bridging and bonding. Bridging social capital pertains to weak tie networks and connects individuals to external resources. It emphasizes the diffusion of information and resources over establishment of emotional support. On the contrary, bonding social capital refers to strong tie networks including family members or close friends with strong personal connections. These links enable them to provide each other with substantive and emotional support (Putnam, 1995).
Social capital is characterized in terms of social network, trust, and reciprocity. In bridging networks people acquire different information from multiple groups, and work for others without expecting any return since they believe they will later benefit. This is called "generalized reciprocity." A macro effect of bridging is increased social tolerance, since people must communicate with people from different backgrounds for future benefits in bridging networks (Miyata, 2007). In bonding networks, people relate closely, so individual relationships create trust and bonding groups create rigid norms reciprocity. In other words, society sanctions people who misbehave to maintain norms.

**Interaction with FB and Social Capital**

Several studies have established a relationship between undergraduate FB use (Ellison, Steinfield & Lampe., 2007; Valenzuela, Park & Kee, 2008) and their social capital. Ellison et al. (2007) found that FB communication practices focused on using the site for social-information seeking purposes were positively associated with online social capital. Aubrey and Rill (2013) reported that using FB for sociability reasons was associated with increased online bridging and bonding. Likewise, FB users reported significantly higher bridging social capital than non-users according to Vitak, Ellison, and Steinfield (2011). Taken together, these studies suggest that users who engage in FB activities are likely to experience gains in online social capital.

However, cross-cultural studies, especially those including Asian countries, are limited. Jiang and de Bruijn (2013) reported that none have addressed the issue of cross-cultural social networking, and conducted their research with a sample of 100 British and 100 Chinese students at the University of Manchester. They found that the greater intensity of cross-cultural FB interactions was positively associated with a perceived increase in cross-cultural social capital. In addition, a follow-up interview study further revealed that the perceived benefits of online social capital depend on both the type of FB interaction and the type of friendship, as well as a combination of situational circumstances and the cultural background of the interviewee.

Jiang, de Bruijn, and De Angeli (2009) found that people from different cultures have different ways of presenting themselves and perceiving others on social networking sites. These studies show that we must pay attention to cultural behaviors for constructing social capital on FB.

Basilisco and Cha (2015) investigated the motives of Filipino FB users and the impact of their usage on social capital and life satisfaction with a sample of 243 participants from a variety of age groups. Among them, 58.4% lived in the Philippines, while 41.6% lived abroad. The identified motivations suggested a significant degree of friendship seeking, entertainment, information, convenience, social capital, and life satisfaction, aside from social support. This study demonstrates Filipinos' unique usage of FB, but this is not the behavior of all undergraduate students.

**Willingness to Communicate and International Posture**

Willingness to Communicate (WTC) measures a person's willingness to initiate communication. It is assumed that WTC is a personality-based, trait-like predisposition relatively consistent across many communication contexts and types of receivers (Richmond & McCroskey, 1992). WTC measurements consist of the four communication contexts of public speaking, talking in meetings, talking in small groups, and talking in dyads, and the three types of receivers of strangers, acquaintance, and friends.

WTC is also used to investigate communication behaviors in relation to second-language learning (Yashima, 2002). Yashima (2002) developed the related notion of "international posture" to explain how learners, such as Japanese students, in contexts lacking meaningful direct contact with target-language speakers manage the motivation to learn a second language. The main characteristics of international posture are described as an "interest in foreign or international affairs, willingness to go overseas to stay or work, readiness to interact with intercultural partners, and, one hopes, openness or a non-ethnocentric attitude towards different cultures" (Yashima, 2002, p. 57). Yashima (2002) found that international posture is a valid construct that relates to motivation to learn and WTC among Japanese university learners of English. The items included in international posture are 1) intergroup approach - avoidance tendency, 2) interest in international vocation or activities, 3) interest in international news, 4) having things to communicate to the world (Yashima, 2009).

Therefore, in our study, WTC and international posture are included to find a relation between social capital and FB, since Filipino and Japanese students must know to switch languages for smooth interactions.

**Problem Statement**

This study focused on perceptions and patterns of FB use among Filipino and Japanese undergraduates, as well as their international postures and WTC, and developed a model to explain the social capital and its antecedents. Specifically, the study sought to answer the following questions:

1. What is the perceived social capital of Filipino and Japanese Facebook users?
2. Is there a relationship between perceived social capital, Facebook usage, international posture, and WTC?

# Research Design and Methods

In this study, an online survey of students with concise instructions in Japanese and English was administered by their professors from June 1st to July 31st, 2016. Although an explanation was given during class, only the FB users in the classes answered after class. Thus, it is a snowball sampling.

The measurement of social capital was adopted from William (2006) and the measurement of WTC and international posture were adopted from Yashima (2009). Regarding the reliability, "The alpha for the full online bridging and bonding scale is.900, and for the offline version, .889 (William, 2006), the alpha for WTC ranges from .85 to well above .90 (McCroskey, 1992), and the alpha for International posture rages from .79 to .80 depending on the subsection (Yahima, 2009). The questionnaire was composed of 1) Demographics, (k = 8); 2) Facebook Usage (k = 39); 3) perceptions of users on FB use (k = 11); 4) Bonding (k = 10); 5) Bridging (k = 10); 6) International posture (k = 14); 7) and WTC (k = 6). We adopted a validated scale to develop our survey questionnaire, employing a five-point scale (Vagias, 2006; 1 = strongly disagree, 5 = strongly agree for the level of agreement and 1 = never, 1 = always for the frequency).

To determine nationality difference, a t-test was performed. In addition, factor analysis was conducted to simplify the data by reducing the observed variables. Confirmatory factor analysis was conducted to identify interrelationships among factors. Lastly, a path analysis based on covariance structure analysis was conducted.

# Results and Discussion

This study obtained 483 valid responses from undergraduate students based on the following: Filipino = 389 (Male = 219, Female = 170, mean of age = 19.21) and Japanese = 94 (Male = 48, Female = 46, mean of age = 20.96); 267 male and 216 female respondents.

**Facebook Usage and Purpose**

FB is commonly used for chatting. Of Filipino respondents, 99.70% used it for text chat, 34.70% for video chat, and 31.60% for voice chat. Of Japanese respondents, 53.19% used it for text chat, 4.26% for video chat, and 2.13% for voice chat. This shows that Japanese students seldom use video and voice chat.
In terms of the purpose of FB use, 92.50% of Filipino respondents to 56.38% of Japanese respondents read news feeds and comments. Of Filipino respondents, 84.30% to 23.40% of Japanese respondents used FB for sending files for non-academic and academic purposes. Of Filipino respondents, 73.80% to 39.36% of Japanese respondents searched for people or friends on FB. Of Filipino respondents, 69.70% watched videos on FB to 37.23% of Japanese respondents. Thus, the usage of sending files by Filipino students was quite different from Japanese.
Figure 1 shows that among 389 Filipino respondents, 32.40% spent less than 5 hours, 9.30% spent less than 4 hours, 16.70% spent less than 3 hours, 17.50% spent less than 2 hours and 15.90% spent less than 1 hour on FB per day. However, as in Figure 1, out of 94 Japanese respondents, 3.19% spent less than 5 hours, 1.06% spent less than 4 hours, 3.19% spent less than 3 hours, 4.26% spent less than 2 hours and 89.36% less than 1 hour on FB in a day.

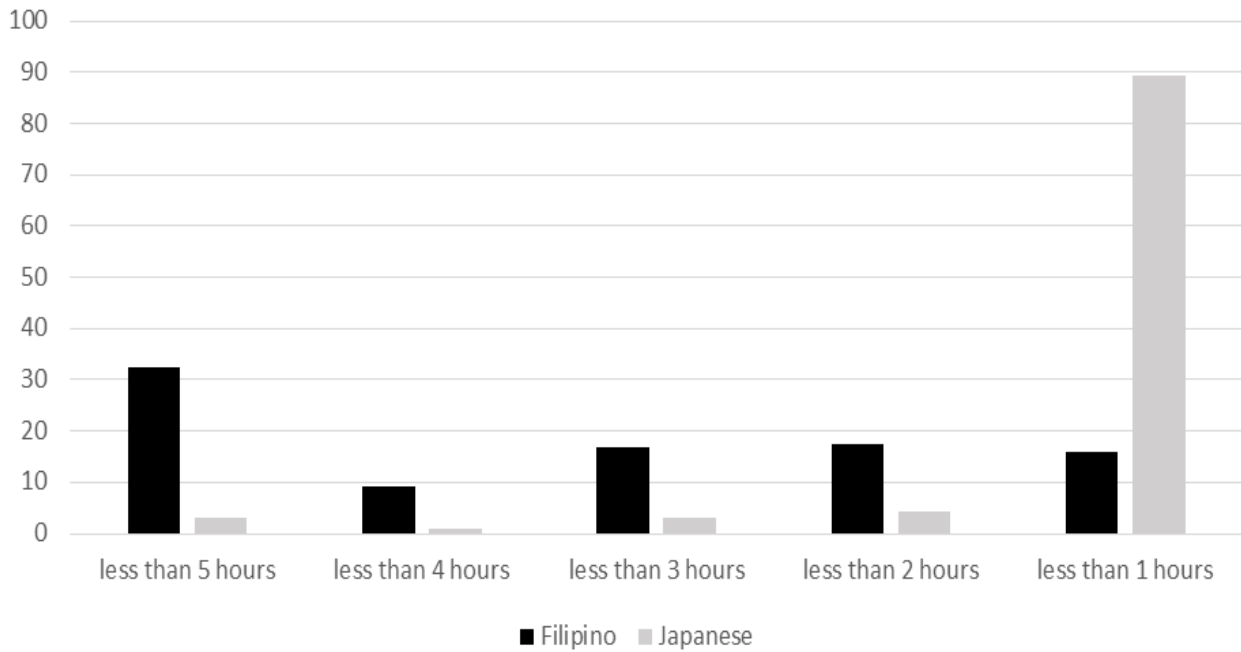

*Figure 1.* Filipino and Japanese Time Spent in Facebook

Moreover, as shown in Figure 2., 65.80% of Filipino respondents were always logged in: 13.90% logged in five times in a day, 6.90% logged in four times, 6.40% logged in three times, 4.40% logged in two times, and 2.60% once a week. Figure 2 shows that 21.28% of Japanese respondents are always online, 26.60% logged in five times, 7.45% logged in four times, 13.83% logged in three times, 7.45% logged in two times, and 18.09% logged in once a week. This shows that Filipino students spend more time on FB than Japanese students.

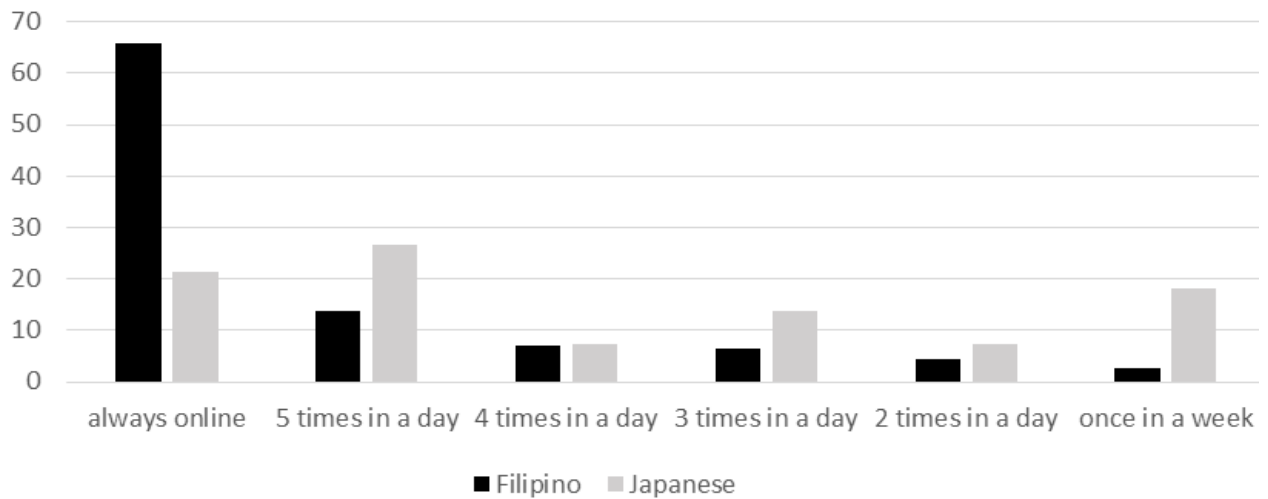

*Figure 2.* Filipino and Japanese Frequency of Facebook Access

Figures 1 and 2 clearly show the different FB-accessing behaviors of Filipino and Japanese students. Typical Filipino students are always online and spend 4 to 5 hours on FB in a day, while Japanese students access it 5 times in a day and spend less than 1 hour online.

Filipino respondents read academic/educational topics (77.40%), social events (71.70%), technology (70.40%), and movies (68.60%), while Japanese respondents read academic/educational topics (45.74%), social events (47.87%), music (37.23%), and rumors/gossip (32.98%). Of Filipino students, 35.70% are interested in religion, to only 1.06% of Japanese students. Thus, Filipino students might use more actively for collaborative learning and gaining knowledge by using FB, compared with Japanese students.

**FB Friends**

Table 1 shows the descriptive statistics and t-test result for Filipino and Japanese students' FB friends. An independent samples t-test was conducted to compare Filipino and Japanese respondents' FB friends, in which there was a significant difference in the number of total FB friends of Filipino (M = 1264.80, SD = 953.94) and Japanese (M = 251.99, SD = 222.57) FB users; t(483) = 18.918, p = 0.000.

Also, there were significant differences in the number of FB friends of Filipino respondents who were their immediate family t(483) = 7.241, p = 0.000; relatives, t(483) = 12.932, p = 0.000; close friends t(483) = 3.989, p = 0.000; classmates t(483) = 10.276, p = 0.000; acquaintances t(483) = 5.155, p = 0.000; strangers t(483) = 12.088, p = 0.000; and foreign friends t(483) = -3.580, p = 0.001.

Table 1
*Descriptive Statistics and T-Test Result*

| FB Friends | Facebook Friends of Filipino and Japanese Students ||||||
|---|---|---|---|---|---|---|
| | Filipino || Japanese || t value | Sig. |
| | Mean | SD | Mean | SD | | |
| Total FB Friends | 1264.8 | 953.94 | 251.99 | 222.57 | 18.918 | 0.000 |
| Immediate Family | 17.9 | 45.79 | 1.01 | 2.14 | 7.24 | 0.000 |
| Relative | 60.86 | 90.28 | 1.49 | 3.43 | 12.93 | 0.000 |
| Close Friends | 108.89 | 215.11 | 48.3 | 102.52 | 3.99 | 0.000 |
| Classmates | 183.34 | 210.46 | 47.39 | 75.82 | 10.28 | 0.000 |
| Acquaintances | 232.8 | 307.15 | 125.14 | 134.93 | 5.16 | 0.000 |
| People They Do Not Know | 423.85 | 646.82 | 22.76 | 48.84 | 12.09 | 0.000 |
| Foreign Friends | 15.7 | 53.7 | 58.99 | 114.22 | -3.58 | 0.001 |

This implies that Filipino respondents tend to be more comfortable and open in accepting FB friends as compared to Japanese counterparts. However, Japanese respondents have more foreign friends compared with Filipino respondents, since Japanese respondents often started using FB specifically to correspond with foreign friends.

**FB Interaction**

Table 2 contains independent sample t-test results comparing FB activities between Filipino and Japanese students. Data show significant differences in FB activities in tagging pictures t(483) = 6.005, p = 0.000, tagging videos t(483) = 12.594, p = 0.000; sharing pictures t(483) = 4.589, p = 0.000 and sharing videos t(483) = 6.814, p = 0.000; liking FB business/commercial page t(483) = 10.465, p = 0.000, FB public figures/fan page t(483) = 6.127, p = 0.000; posting pictures t(483) = 4.634, p = 0.000 and posting videos t(483) = 7.684, p = 0.000; and using emojis and GIFs in posting t(483) = 11.711, p = 0.000 and chatting t(483) = 14.436, p = 0.000. The data imply that Filipino students more actively tag, share, like, and post using emojis and GIFs as compared to Japanese students. The highest percentage of Filipino students posted pictures with friends, as well as Japanese respondents, at 83.30% and 38.30%, respectively. However, only 5.32% of Japanese respondents, compared to 76.10% of Filipino respondents, posted family pictures on FB.

There is a very notable similarity between Japanese and Filipino students in liking posts t(483) = -0.787, p = .433.

Table 2
*T-Test Result of Filipino and Japanese students FB Activity*

| | Activity | Facebook Interaction of Filipino and Japanese Students | |
|---|---|---|---|
| | | t value | Sig. |
| TAGGING | Pictures | 6.005 | .000 |
| | Videos | 12.549 | .000 |
| SHARING | Pictures | 4.589 | .000 |
| | Videos | 6.814 | .000 |
| LIKING | Facebook Business/Commercial Page | 10.465 | .000 |
| | Facebook Public Figures/Fan Page | 6.127 | .000 |
| | Facebook Post (Comments, Pictures and Videos) | -0.787 | .433 |
| POSTING | Pictures | 4.634 | .000 |
| | Videos | 7.684 | .000 |
| USE OF EMOJI AND GIF | Posting | 11.711 | .000 |
| | Chatting | 14.436 | .000 |

Confirmatory factor analysis was conducted to assess the convergent and discriminant validities of the survey instrument (Rattray & Jones, 2007) used in the structure equation modeling. The following items were drawn through exploratory factor analysis as shown in Table 3. This corresponds to the six factors from the latent variables: International Posture, WTC, Perception, Bridging, Bonding, and Utilization. The same model was used in the covariance analysis of structure. The data for each group were loaded in the developed model.

Table 3
*Factor Analysis of Items*

| Items | Factors | | | | | |
|---|---|---|---|---|---|---|
| | 1 | 2 | 3 | 4 | 5 | 6 |
| I often talk about situations and events in foreign countries with my family and/or friends. | **0.876** | 0.051 | 0.023 | -0.051 | 0.028 | -0.042 |
| I often read and watch news about foreign countries. | **0.832** | -0.071 | -0.047 | -0.043 | 0.059 | 0.106 |
| I have a strong interest in international affairs. | **0.854** | 0.037 | 0.037 | 0.108 | -0.078 | -0.089 |
| Talking to a small group of strangers. | 0.008 | **0.824** | -0.006 | -0.028 | 0.011 | 0.012 |
| Talking with a salesperson in a store. | 0.002 | **0.831** | -0.003 | 0.023 | 0.003 | 0.012 |
| Facebook is part of my everyday activity. | 0.014 | -0.036 | **0.854** | -0.086 | 0.014 | 0.021 |
| I am proud to tell people I am on Facebook. | -0.009 | 0.007 | **0.748** | 0.074 | -0.010 | 0.008 |
| Interacting with people on Facebook gives me new people to talk to. | -0.038 | 0.024 | -0.061 | **0.910** | -0.071 | 0.042 |
| Interacting with people on Facebook makes me feel connected to the bigger picture. | 0.042 | -0.039 | 0.057 | **0.834** | 0.119 | -0.024 |
| The people I interact on Facebook who would put their reputation on the line for me. | 0.031 | -0.037 | -0.06 | -0.011 | **0.866** | 0.074 |
| The people I interact with on Facebook would share their last money with me. | -0.05 | 0.047 | 0.053 | 0.034 | **0.827** | -0.087 |
| I have used Facebook to check out someone I met socially. | -0.025 | 0.011 | 0.022 | -0.08 | 0.042 | **0.792** |
| I use Facebook to learn more about other people in my classes. | 0.017 | 0.016 | 0.006 | 0.13 | -0.06 | **0.776** |
| Factors | 1 | 2 | 3 | 4 | 5 | 6 |
| 1 | 1 | 0.108 | 0.079 | 0.268 | 0.094 | 0.152 |
| 2 | | 1 | 0.023 | 0.176 | 0.207 | 0.085 |
| 3 | | | 1 | 0.188 | 0.254 | 0.276 |
| 4 | | | | 1 | 0.356 | 0.363 |
| 5 | | | | | 1 | 0.215 |
| 6 | | | | | | 1 |

A path analysis was conducted to determine each construct's path coefficients and determine factors' interrelationship. There were thirteen (13) observed variables or items and six (6) constructs. Data between the Filipino respondents in Figure 3 and the Japanese respondents in Figure 4 were grouped using the same model and used for analysis.

The model-fit indices technique was used to assess how the priori model fit the data. The hypothesized model shows a good fit with GFI (Good of Fit Index) = .916, AGFI (Adjusted Goodness of Fit Index) = .870, CFI (Comparative Fit Index) = .918, IFI (Incremental Fit Index) = .919, TLI (Tucker Fit Index) = .891, and RMSEA (Root mean Square Error of Approximation) = .056. The value $0.7 \leq x \leq 0.9$ for AGFI and TLI is an acceptable fit.

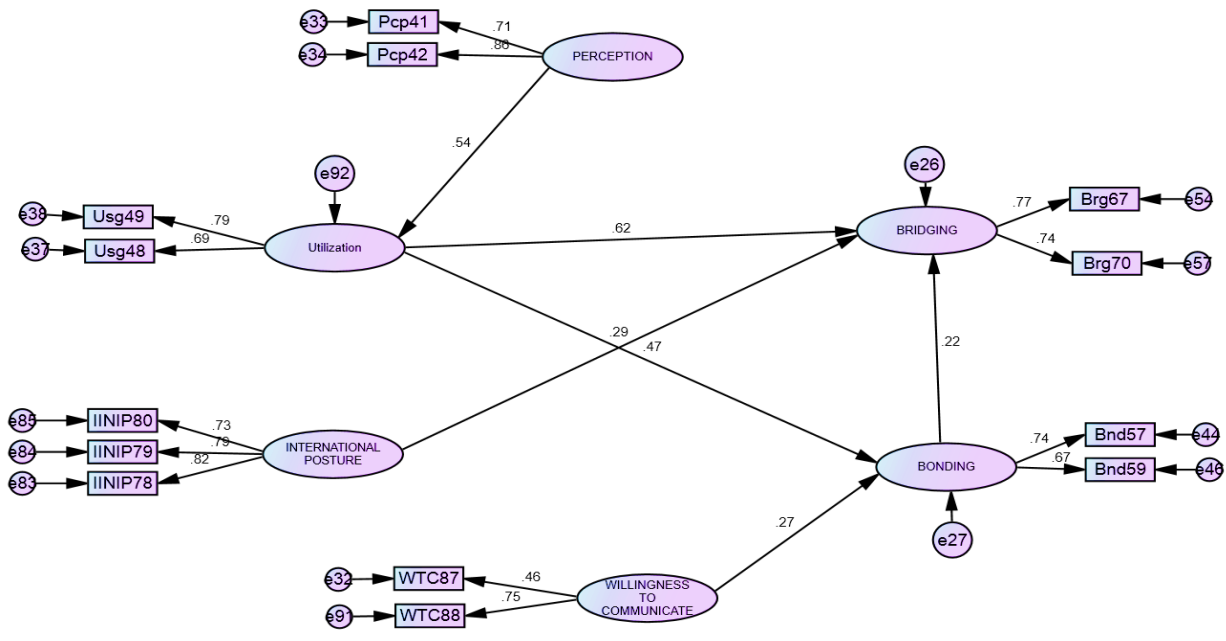

*Figure 3.* Filipino Interrelation Model

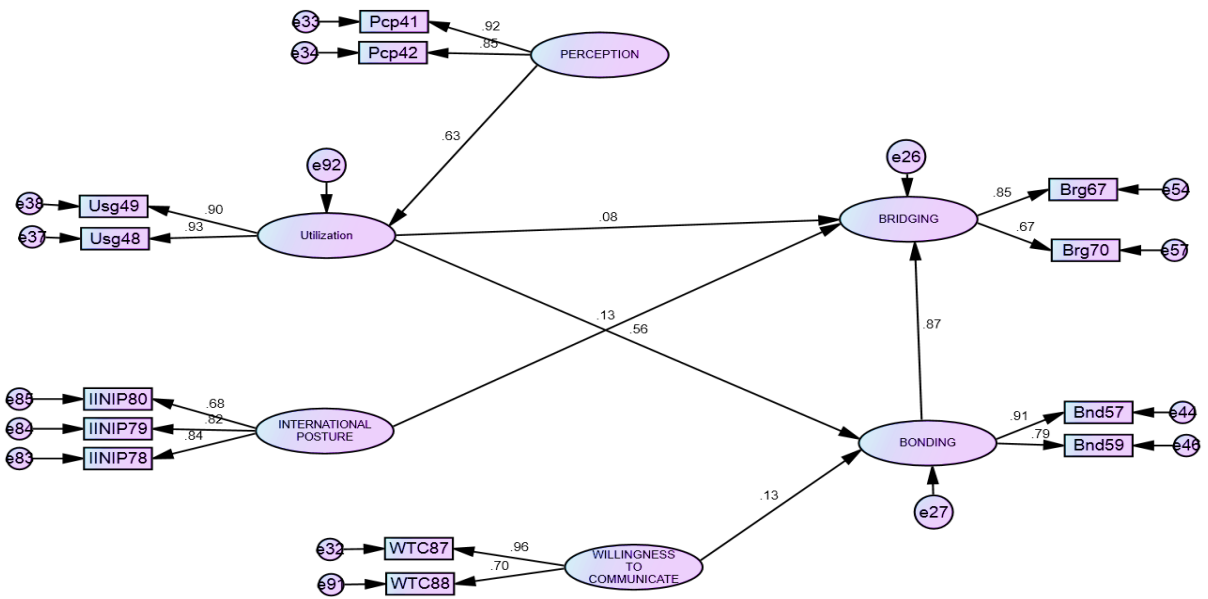

*Figure 4.* Japanese Interrelation Model

As in Table 4, the path coefficient of Filipino students showed a significant effect among constructs such as Perception to Utilization (β = .538, p < .01), WTC to Bonding (β = .266, p <. 05), Utilization to Bonding (β = .472, p < .01), Utilization to Bridging (β = .624, p < .01), Bonding to Bridging (β = .222, p < .01), International Posture to Bridging (β = .288, p < .01). The path coefficient of Japanese students showed a significant effect among constructs such as Perception to Utilization (β = .634, p < .01), Utilization to Bonding (β = .558, p < .01), and Bonding to Bridging (β = .872, p < .01).

Table 4
*The Path Coefficient of Filipino and Japanese Students*

| Path of Constructs | | | Filipino Students n = 389 | | Japanese Students n = 94 | |
|---|---|---|---|---|---|---|
| | | | Path Coefficient | P | Path Coefficient | P |
| Utilization | <--- | Perception | 0.538 | .000 | 0.634 | .000 |
| Bonding | <--- | Willingness To Communicate | 0.266 | 0.038 | 0.133 | 0.177 |
| Bonding | <--- | Utilization | 0.472 | .000 | 0.558 | .000 |
| Bridging | <--- | Utilization | 0.624 | .000 | 0.083 | 0.427 |
| Bridging | <--- | Bonding | 0.222 | 0.003 | 0.872 | .000 |
| Bridging | <--- | International Posture | 0.288 | .000 | 0.13 | 0.113 |

Notes: *** p-value < 0.01; ** p-value < 0.05; * p-value < 0.10

The Filipino interrelation model suggests that perception affected utilization, which in turn affected bridging and bonding. Bonding also affected bridging. International posture affected bridging, while WTC affected bonding. Thus, Filipino students' utilization of FB directly related to bridging and boding, and accumulated bonding capitals also affected bridging. To enhance bridging for Filipino students, it is advisable to enhance international posture, specifically "interests in international news" (Yashima, 2009). Additionally, bonding mediates WTC and bridging, so we can expect Filipino students who do not mind talking to salespersons or in front of a small group of strangers to gain bonding capital and turn it into bridging.

However, the interrelation model suggests that Japanese students' perception affected utilization, which, in turn affected only bonding. Then, bonding affected bridging. In other words, Japanese students must enhance bonding capital first, and mediate it to enhance bridging capital. Their international posture and WTC did not relate to bridging or bonding social capital. Japanese students seem not to challenge to search for new foreign friends, thus, thier "inward tendencies" are revealed even on FB sites.

If Filipino and Japanese students meet face-to-face and make a strong tie, they gain bonding capital by interacting on FB and turn it into bridging capital. Filipino students with WTC might guide Japanese students to gain bonding capital, leading to enhanced bridging capital.

# Conclusion

The purpose of this study is to investigate perception and patterns of FB among Filipino and Japanese undergraduate students and the relationship of these factors in creating and maintaining students' social capital. FB was selected among other SNS because it easily creates the sites of online intercultural exchange between Filipino and Japanese students for educational purpose.

The study clearly revealed Filipino and Japanese undergraduate students' characteristic uses of FB. Filipino students had five times as many as friends than Japanese students. They were friends not only with close friends, immediate family, and relatives, but also people they did not know. However, Japanese students did not include family or people they did not know, but did include foreign friends. This means Japanese students intentionally select FB to keep contacts with foreign friends they once made, and they use other SNS such as LINE to contact their family or immediate friends.

For Filipino students, it is very common to have family or relatives living abroad. Thus, FB is very convenient and less expensive tool to contact with them. Filipino students might use FB because of its convenient feature, this is about the same motivation as the results given by Basilisco and Cha (2015).

According to Ellison et al. (2007), "we might expect Facebook usage to have less of an impact on bonding than bridging social capital given the affordance of this service. It can lower barriers to participation and therefore may encourage the formation of weak ties but not necessarily create the close kinds of relationships that are associated with bonding social capital"(p.28). In our study, the Japanese and Filipino students show the similar behaviors in terms of liking posts of comments, pictures, and videos. Thus, at least, it can be said that the function of "liking" might be the easiest way to interact with friends. However, this feature does not create bridging capital automatically. Our interrelation model shows the differences between Filipino students and Japanese students towards the usage of FB in relation to social capital. Filipino students use FB actively however, their social capital are constructed among Filipino people.

Japanese students have more foreign friends than Filipino. They use FB in order to strengthen their friendships, and not for searching a new friend, since internal posture and WTC did not affect any social capital.

Therefore, when we think "virtual internationalization" among Filipino and Japanese students, it is advisable to cultivate the bonding capital first and then it will in turn have an effect on bridging capital. Bridging capital is the key to cultivate the students' tolerance towards ambiguity, trustworthiness and generalized reciprocity, and is necessary to enhance intercultural competences. Weisinger and Salipante (2005) pointed that it is necessary to create a bonding capital first, and then, to create a bridging capital for maintaining mulita dimensional aspects of the volunteer association. However, in our study, two interrelation models only show the different purpose of FB use by Filipino students and Japanese students, and could not show how to combine the bonding capital and the bridging capital for the intercultural interaction. That will be our future task.

Another limitation of our study is the number of Japanese respondents. We had difficulties of finding FB users among Japanese undergraduate students due to the rapid spread of LINE in Japan. We need to consider the rapid progress of SNS for educational use.